\documentclass{ws-procs9x6}

\def\cal#1{\mathcal#1}

\def\ga{\gamma}
\def\de{\delta}
\def\ep{\epsilon}

\def\et{\eta}

\def\la{\lambda}

\def\si{\sigma}

\def\ta{\tau}

\def\vp{\varphi}
\def\ch{\chi}

\def\om{\omega}
\def\Ga{\Gamma}
\def\De{\Delta}
\def\Th{\Theta}

\def\Ph{\Phi}

\def\mn{{\mu\nu}}

\def\cl{{\cal L}}

\def\fr#1#2{{{#1} \over {#2}}}
\def\frac#1#2{\textstyle{{{#1} \over {#2}}}}
\def\pt#1{\phantom{#1}}
\def\prt{\partial}
\def\vev#1{\langle {#1}\rangle}

\def\half{{\textstyle{1\over 2}}}
\def\lsim{\mathrel{\rlap{\lower4pt\hbox{\hskip1pt$\sim$}}
    \raise1pt\hbox{$<$}}}
\def\gsim{\mathrel{\rlap{\lower4pt\hbox{\hskip1pt$\sim$}}
    \raise1pt\hbox{$>$}}}

\def\etal {{\it et al.}}
\newcommand{\beq}{\begin{equation}}
\newcommand{\eeq}{\end{equation}}
\newcommand{\bea}{\begin{eqnarray}}
\newcommand{\eea}{\end{eqnarray}}
\newcommand{\bse}{\begin{subequations}}
\newcommand{\ese}{\end{subequations}}
\newcommand{\rf}[1]{(\ref{#1})}

\def\to{\rightarrow}

\def\mix{\leftrightarrow}

\def\nub{\bar\nu}
\def\vp{\vec p}
\def\cmat{{\cal C}}

\def\heff{h_{\rm eff}}
\def\Ueff{U_{\rm eff}}
\def\tu{\widetilde U}
\def\bu{\overline U}

\def\mt{\widetilde m^2}

\def\ring#1{{\mathaccent'27 #1}}
\def\cri{\ring{c}}

\def\a3em{\check{a}}
\def\cee{\cri}
\def\Dtm{\De m^2_{\Th}}
\def\Dtmz{\De m^2_{0^\circ}}

\begin{document}

\title{Neutrino Oscillations and Lorentz Violation}

\author{V.\ Alan Kosteleck\'y}
\address{Physics Department, Indiana University, 
         Bloomington, IN 47405, U.S.A.}
%{IUHET 464, January 2004} 

\maketitle

\abstracts{
Existing experimental data for neutrino oscillations
are consistent with Lorentz-violating massless neutrinos.
This talk summarizes some aspects of neutrino oscillations 
from the perspective of Lorentz and CPT violation 
in effective quantum field theory.
}

\section{Introduction}

Most data acquired to date by experiments in particle physics 
are successfully described by the minimal Standard Model (SM).
One exception is the observation of neutrino oscillations,
which are now well established and provide a compelling indication 
that the minimal SM requires modification.
The inclusion of small masses in the neutrino sector of the SM
suffices to describe the bulk of the experimental results, 
although a definitive understanding 
awaits the results of ongoing and future measurements.
From a theoretical perspective,
these modifications to the SM could arise 
as suppressed low-energy energy effects from 
an underlying unified theory at the Planck scale,
$m_P \simeq 10^{19}$ GeV.
This suggests the interesting possibility that
neutrino oscillations might represent signals
from the Planck scale.
 
This talk considers the description of neutrino oscillations
in terms of a promising candidate Planck-scale signal,
namely, 
violations of relativity arising through Lorentz and CPT 
breaking.\cite{cpt01}
The focus is on neutrino behavior 
from the perspective of Lorentz and CPT violation 
in effective quantum field theory.
The general effective field theory
describing Lorentz and CPT violation
is called the Standard-Model Extension\cite{ck,ak}
(SME)
and was the basis for the original suggestion that 
Lorentz and CPT violation might occur in neutrinos.\cite{ck}
In the intervening years,
several theoretical investigations of this possibility
within the context of the SME have been 
performed,\cite{fc1,fc2,fc3,fc4,fc5,nu,nu2,chki}
mostly for special models involving
only a few coefficients for Lorentz and CPT violation.
Here, 
a comprehensive theoretical study of Lorentz and CPT violation
in neutrinos is outlined
and a potentiallly realistic model is described.
Details about the ideas discussed here 
can be found in the literature.\cite{nu,nu2} 

\section{Theory}\label{theory}

The lagrangian of the SME consists of the SM 
and gravitational lagrangian
supplemented by all possible coordinate-invariant terms
constructed with SM and gravitational fields
and violating Lorentz symmetry.\cite{ak}
The extra terms involve Lorentz-violating operators
coupled to coefficients with Lorentz indices.
Various origins have been suggested for these operators.
A widely applicable and theoretically attractive source 
is physical spontaneous Lorentz violation,
first suggested in gravitationally coupled field theories 
and string theory\cite{ks}
and then extended to incorporate CPT violation.\cite{kp}
Proposed sources also include  
noncommutative field theory,\cite{ncqed}
non-string approaches to quantum gravity,\cite{qg}
random dynamics,\cite{fn}
and multiverses.\cite{bj}
Measurements of SME coefficients for Lorentz violation 
have attained Planck-scale sensitivity
in experiments with
mesons,\cite{hadronexpt,kpo,hadronth}
baryons,\cite{ccexpt,spaceexpt,cane}
electrons,\cite{eexpt,eexpt2}
photons,\cite{photonexpt,photonth,cavexpt,km}
and muons.\cite{muons}

This part of the talk
considers the basic theory of neutrino behavior
in the presence of Lorentz and CPT violation
in Minkowski spacetime.
The main focus is oscillation phenomena,\cite{pdg}
but the formalism described can also be applied
to other classes of experiments such as  
supernova studies,
direct mass searches,
and tests for neutrinoless double-beta decay.

\subsection{Effective hamiltonian}

Consider a general theory for $N$ neutrino species,
incorporating all Majorana- and Dirac-type couplings 
of left- and right-handed neutrinos,
both with and without Lorentz violation. 
Denote the neutrino fields by the Dirac spinors
\{$\nu_e$, $\nu_\mu$, $\nu_\ta$,\ldots \}
and their charge conjugates 
\{$\nu_{e^C}\equiv \nu^C_e$,
$\nu_{\mu^C}\equiv \nu^C_\mu$,
$\nu_{\ta^C}\equiv \nu^C_\ta$,\ldots \},
where $\nu^C_a \equiv C\nub^T_a$ as usual.
Collect
all fields and conjugates into a quantity $\nu_A$,
where $A$ ranges over the $2N$ possibilities
$\{e,\mu,\ta,\ldots, e^C,\mu^C,\ta^C,\ldots \}$,
so that charge conjugation becomes 
a linear transformation $\nu^C_A = \cmat_{AB} \nu_B$.
In this notation,
the dominant contributions to the 
general equations of motion for free propagation
yield the first-order differential equation
\beq
(i\Ga^\nu_{AB}\prt_\nu-M_{AB})\nu_B=0 ,
\label{de}
\eeq
extending the usual equations of motion 
for Dirac and Majorana neutrinos.

In Equation \rf{de},
$\Ga^\mu_{AB}$, $M_{AB}$ are constant matrices
in spinor space.
They can be decomposed as
\bea
\Ga^\nu_{AB} &\equiv&
\ga^\nu \de_{AB}
+ c^\mn_{AB}\ga_\mu
+ d^\mn_{AB}\ga_5\ga_\mu
+ e^\nu_{AB}
+ if^\nu_{AB}\ga_5
+ \half g^{\la\mn}_{AB}\si_{\la\mu} ,
\nonumber\\
M_{AB} &\equiv&
m_{AB}
+im_{5AB}\ga_5
+ a^\mu_{AB}\ga_\mu
+ b^\mu_{AB}\ga_5\ga_\mu
+ \half H^\mn_{AB}\si_\mn .
\label{GaM}\eea
Here,
$m$ and $m_5$ are mass terms.
All other coefficients control terms 
violating Lorentz symmetry,
and the coefficients $a$, $b$, $e$, $f$, $g$
control CPT violation as well.
The matrices $\Ga^\mu_{AB}$, $M_{AB}$ 
must satisfy certain conditions following from 
hermiticity of the theory 
and from the interdependence of $\nu$ and $\nu^C$.
For example,
all the coefficients for Lorentz and CPT violation 
in Eq.\ \rf{GaM}
are hermitian in generation space.

Obtaining the hamiltonian associated with Eq.\ \rf{de} 
requires handling the unconventional coefficient 
for the time derivative,
which can be done following established procedures.\cite{kle}
For simplicity,
consider here only the minimal case with $N=3$
and a standard seesaw mechanism\cite{seesaw}
suppressing the propagation of right-handed neutrinos,
and restrict attention to leading-order Lorentz and CPT violation.
Amplitudes $b_{e,\mu,\ta}(t;\vp)$ and $d_{e,\mu,\ta}(t;\vp)$ 
can be defined that approximate active neutrinos
and active antineutrinos of momentum $\vp$.
A somewhat lengthy calculation\cite{nu2}
establishes that the time evolution 
of these amplitudes is 
\beq
\left(
\begin{array}{c}
b_a(t;\vp)\\
d_a(t;\vp)
\end{array}
\right)
=\exp(-i\heff t)_{ab}
\left(
\begin{array}{c}
b_b(0;\vp)\\
d_b(0;\vp)
\end{array}
\right) ,
\label{Ut}
\eeq
where $\heff$ is the effective hamiltonian
for flavor neutrino propagation,
given at leading order by
\bea
&
\hskip-110pt
(\heff)_{ab}=
|\vp|\de_{ab}
\left(\begin{array}{cc}
1 & 0 \\
0 & 1
\end{array}\right)
+\fr{1}{2|\vp|}
\left(\begin{array}{cc}
(\mt)_{ab} & 0 \\
0& (\mt)^*_{ab}
\end{array}\right)
&
\nonumber \\
&
+\fr{1}{|\vp|}
\left(\begin{array}{cc}
\hskip -10pt
[(a_L)^\mu p_\mu-(c_L)^\mn p_\mu p_\nu]_{ab} &
\hskip -10pt
-i\sqrt{2} p_\mu (\ep_+)_\nu
[(g^{\mn\si}p_\si-H^\mn)\cmat]_{ab} \\
i\sqrt{2} p_\mu (\ep_+)^*_\nu
[(g^{\mn\si}p_\si+H^\mn)\cmat]^*_{ab} &
[-(a_L)^\mu p_\mu-(c_L)^\mn p_\mu p_\nu]^*_{ab}
\end{array}\right) . 
&
\nonumber\\
\label{heff}
\eea
In this equation,
$(c_L)^\mn_{ab}\equiv(c+d)^\mn_{ab}$ and
$(a_L)^\mu_{ab}\equiv(a+b)^\mu_{ab}$,
$p_\mu=(|\vp|;-\vp)$ at leading order,
and the complex vector $(\ep_+)_\mu$ can be chosen as 
$(\ep_+)^\nu=\frac{1}{\sqrt{2}}(0;\hat\ep_1+i\hat\ep_2)$,
where $\hat\ep_1$, $\hat\ep_2$ are real
and $\{ \vp/|\vp|, \hat\ep_1, \hat\ep_2 \}$
form a right-handed orthonormal triad.

The first line of the above equation
involves the diagonal kinetic term of the minimal SM
and the usual Lorentz-conserving neutrino-mass term.
In the second line,
the coefficients $(a_L)^\mu_{ab}$, $(c_L)^\mn_{ab}$
determine the dominant Lorentz-violating effects 
in neutrino-neutrino mixing,
while $(g^{\mn\si}\cmat)_{ab}$, $(H^\mn\cmat)_{ab}$
generate Lorentz-violating neutrino-antineutrino mixing.
The former two types of coefficient preserve 
SU(3)$\times$SU(2)$\times$U(1) 
and appear in the minimal SME,
while the latter two involve Majorana-type couplings 
breaking both SU(3)$\times$SU(2)$\times$U(1) 
and lepton number.
Most of these coefficients lead to physical effects,
but a few combinations are unphysical due to symmetries 
or to the existence of field redefinitions
that can eliminate them.\cite{ck,kle,cm,bek}
Note that the properties of the coefficients for Lorentz violation
in Eq.\ \rf{GaM}
imply that the CPT-conjugate hamiltonian is obtained
by changing the sign of the coefficients $a_L$ and $g$.
Note also that none of the terms in Eq.\ \rf{heff}
correspond to independent mass matrices
for neutrinos and antineutrinos 
as recently proposed,\cite{mmb}
a result consistent with Greenberg's formal 
impossibility proof.\cite{owg}

The effective hamiltonian \rf{heff}
describes all contributions 
from operators of renormalizable dimension
and therefore provides a definitive foundation 
for the treatment of Lorentz and CPT violation in neutrinos. 
Operators of nonrenormalizable mass dimension $n>4$ 
could also be significant,
depending partly on the degree to which they are suppressed
by powers of the Planck scale.\cite{kpo}
In fact,
at energies well beyond those relevant for current experiments,
Lorentz-violating terms of nonrenormalizable dimension 
may be necessary for stability and causality.\cite{kle}
Some consequences of these more general operators are
discussed in the literature.\cite{nu,nu2,bef}

The generality of the effective hamiltonian \rf{heff}
implies that it must also contain the effects of matter interactions.
Indeed,
the effective lagrangian for neutrino propagation
in normal matter is 
$\De\cl_{\rm matter}=
 -\sqrt{2}G_Fn_e\nub_e\ga^0P_L\nu_e
+(G_Fn_n/\sqrt{2})\nub_a\ga^0P_L\nu_a$,
which means that matter effects are incorporated
by coefficient contributions of the form 
$(a_{L,\rm eff})^0_{ee}=G_F(2n_e-n_n)/\sqrt{2}$
and $(a_{L,\rm eff})^0_{\mu\mu}=(a_{L,\rm eff})^0_{\ta\ta}=
-G_Fn_n/\sqrt{2}$,
where $n_e$ and $n_n$ are the number
densities of electrons and neutrons.

\subsection{Neutrino mixing}

To extract the implications for neutrino mixing, 
it is useful to diagonalize the effective hamiltonian.
This involves a $6\times 6$ unitary matrix $\Ueff$,
$\heff =  \Ueff^\dag E_{\rm eff} \Ueff$,
where $E_{\rm eff}$ is a $6\times 6$ diagonal matrix.
There are therefore as many as six independent propagating
states excluding sterile neutrinos,
five possible eigenvalue differences,
and hence five independent oscillation lengths
in Lorentz-violating mixing.
In contrast,
the Lorentz-covariant case allows only three 
independent propagating states and 
two independent oscillation lengths.
Note also that the five eigenvalue differences are 
{\it not} normal mass differences
since their energy dependences are unconventional.

Denote the six independent propagating states 
by the amplitudes $B_J(t;\vp)$, $J=1,\ldots ,6$. 
Then,
$B_J(t;\vp)=\tu_{Ja}b_a(t;\vp)+\bu_{Ja}d_a(t;\vp)$,
where $\Ueff$ has been separated into $6\times 3$ matrices
$\Ueff=(\tu, \bu)$.
The time evolution operator $S_{ab}(t)$ becomes 
\beq
S_{ab}(t)
= (\Ueff^\dag e^{-iE_{\rm eff}t} \Ueff)_{ab}
= \left(\begin{array}{cc}
S_{\nu_a\nu_b}(t) & S_{\nu_a\nub_b}(t)\\
S_{\nub_a\nu_b}(t) & S_{\nub_a\nub_b}(t)
\end{array}\right) ,
\label{Ut2}\eeq
where $E_{(J)}$ are the diagonal values of $E_{\rm eff}$.
This expression yields the oscillation probabilities
at time $t$.
Thus,
the probability for a neutrino of type $b$
to oscillate into a neutrino of type $a$ is 
$P_{\nu_b\to\nu_a}(t)=|S_{\nu_a\nu_b}(t)|^2$,
while that for a neutrino of type $b$
to oscillate into an antineutrino of type $a$ is 
$P_{\nu_b\to\nub_a}(t)=|S_{\nub_a\nu_b}(t)|^2$.
Similarly, 
for antineutrinos we have
$P_{\nub_b\to\nu_a}(t)=|S_{\nu_a\nub_b}(t)|^2$
or 
$P_{\nub_b\to\nub_a}(t)=|S_{\nub_a\nub_b}(t)|^2$.

The CPT properties of the transition amplitudes are
$S_{\nu_a\nu_b}(t)
\longleftrightarrow
S^*_{\nub_a\nub_b}(-t)$
and
$S_{\nub_a\nu_b}(t)
\longleftrightarrow
-S^*_{\nu_a\nub_b}(-t)$.
If CPT is unbroken these relations become equalities,
whereupon the first generates the standard result
that CPT invariance implies
$P_{\nu_b\to\nu_a}(t) = P_{\nub_a\to\nub_b}(t)$,
while the second implies
$P_{\nu_b\rightleftarrows\nub_a}(t) = 
P_{\nu_a\rightleftarrows\nub_b}(t)$.
However, 
negation of terms in these results can fail.
For example,
CPT violation need not imply 
$P_{\nu_b\to\nu_a}(t) \neq P_{\nub_a\to\nub_b}(t)$.

For the above analysis of the effective hamiltonian
and transition probabilities,
the choice of observer reference frame is irrelevant 
because the physics is coordinate invariant
and in particular is observer Lorentz invariant.\cite{ck}
However,
since particle Lorentz symmetry is violated,\cite{ck}
it is advisable to adopt a standard frame
to report experimental measurements 
of the coefficients for Lorentz violation.
Conventionally,
this is taken to be a Sun-centered celestial equatorial frame 
with coordinates $(T,X,Y,Z)$.
The $Z$ direction is aligned with the Earth's rotational axis,
and the $X$ direction points towards the vernal equinox.
The coefficients for Lorentz violation
in any inertial frame can be related 
to those in the standard Sun-centered frame 
by an appropriate observer Lorentz transformation.\cite{km}
Since neutrino oscillations in the presence of Lorentz violation
can exhibit orientation-dependent effects,
it is also convenient to define 
a standard parametrization in the Sun-centered frame
for the direction of neutrino propagation $\hat p$ 
and for the $\hat\ep_1$, $\hat\ep_2$ vectors 
introduced above:
\bea
\hat p&=&(\sin\Th\cos\Ph,\sin\Th\sin\Ph,\cos\Th) ,\nonumber\\
\hat\ep_1&=&(\cos\Th\cos\Ph,\cos\Th\sin\Ph,-\sin\Th) , \nonumber\\
\hat\ep_2&=&(-\sin\Ph,\cos\Ph,0) .
\label{vectors}
\eea
Here,
$\Th$ and $\Ph$ are the celestial colatitude 
and longitude of propagation,
respectively.
This parametrization is assumed in what follows.

\section{Sensitivities}\label{sens}

No convincing experimental evidence for Lorentz violation
presently exists.
The size of theoretically predicted effects varies,
but it is reasonable to expect\cite{kp}
that observables for Lorentz and CPT violation 
are suppressed by some power of the dimensionless ratio 
$r = m/m_P \lsim 10^{-17}$,
where $m$ is the relevant low-energy scale
and $m_P$ is the Planck mass.
Note that the physics governing experiments
on neutrino oscillations is controlled
by the dimensionless ratio $r_\nu = \De m^2/E^2$, 
where $\De m^2 \lsim 10^{-20}$ GeV
and $10^{-4}$ GeV $< E < 10^3$ GeV.
The ratios $r$ and $r_\nu$ are similar in range,
suggesting that the natural scale for Lorentz violation 
could be comparable to the natural scale of neutrino oscillations.

Sensitivities to certain coefficients for Lorentz violation
in the fermion and photon sectors 
have now reached parts in $10^{-30}$ or better.
It might therefore seem plausible that theoretical considerations
involving symmetries or radiative corrections
could generate constraints on Lorentz violation in neutrinos 
from these impressive sensitivities in other sectors, 
but this idea largely fails in practice.
At tree level,
leading-order perturbative calculations 
involve only flavor-diagonal coefficients,
and the Lorentz-violating 
charged-lepton and neutrino sectors
are completely independent of these.
Also,
although radiative corrections can constrain 
a few neutrino effects under favorable circumstances,\cite{fc5} 
the properties of the electroweak sector 
typically ensure the independence of the charged- 
and neutral-lepton sectors 
at leading order in Lorentz violation 
even under radiative corrections.\cite{nu2}

Various types of experiment can be sensitive
to Lorentz- and CPT-violating effects,
depending on the neutrino behavior studied.
For example,
the energy dependence of Lorentz-violating oscillations 
can be unconventional. 
Recall that the physically relevant dimensionless combination 
controlling neutrino oscillations induced by  
mass-squared differences $\De m^2$
is $\De m^2L/E$,
where $L$ is the baseline distance and $E$ is the neutrino energy.
In contrast,
oscillations induced by Lorentz violation
are controlled by the dimensionless combinations
$a^\mu L$, $b^\mu L$, $H^\mn L$
and $c^\mn LE$, $d^\mn LE$, $g^{\mn\si} LE$,
as readily seen from Eq.\ \rf{heff}.
In general,
oscillations generated by a coefficient $(k_d)^{\la\ldots}$
for a Lorentz-violating operator
of nonrenormalizable dimension $n=d+3$
are controlled by $(k_d)^{\la\ldots} LE^d$.

Another unconventional effect is direction-dependence
of neutrino behavior,
which results from violation of rotation symmetry.\cite{nu2}
This has implications both 
for comparisons between different experiments
and for the analysis of experiments involving multiple sources.
The orientation of the neutrino beam 
and the location of the source relative to the detector
can affect neutrino oscillations.
The daily rotation of the Earth induces 
apparent periodic changes of the coefficients for Lorentz violation
in the laboratory
that would appear as time variations in oscillation data
at multiples of the sidereal frequency
$\om_\oplus\simeq 2\pi/$(23 h 56 min).
Also,
observable annual variations in the solar-neutrino flux 
could arise from the change in the location of the detector
as the Earth orbits the Sun.

Lorentz violation can also lead to novel resonance effects 
in neutrino oscillations.
The usual MSW resonances\cite{msw}
occur when neutrino interactions with local matter
become comparable to effects from mass,
which alters the structure of the effective hamiltonian.
Many other types of resonances can be induced by Lorentz violation,
involving various combinations of 
coefficients for Lorentz violation, masses, and matter effects.
For instance, 
resonances are possible with neither mass nor matter
that are triggered by distinct coefficients for Lorentz violation.  
Known explicit examples include 
a two-generation vacuum resonance 
involving a single nonzero coefficient $(a_L)^T$ 
and a mass term,\cite{fc2}
and a three-generation orientation-dependent vacuum resonance 
involving two coefficients $(a_L)^Z$ and $(c_L)^{TT}$ 
without mass.\cite{nu}

Certain experimental signals can be regarded as characteristic
of Lorentz violation in neutrino oscillations. 
Six model-independent classes of signal
exist that would offer evidence for Lorentz violation.\cite{nu2}
{\it Spectral anomalies}
can arise because each coefficient for Lorentz violation
introduces unconventional energy dependence.
These can generate complicated energy dependences
in both oscillation lengths and mixing angles.
{\it $L$--$E$ conflicts}
represent any null or positive measurement
in a region of $L$--$E$ space that conflicts with
all mass-based scenarios.
Of the six classes mentioned here,
only this one presently has some positive evidence,
the LSND anomaly.\cite{lsnd}
{\it Periodic variations}
are induced by rotation-symmetry breaking
and include both sidereal and annual variations.
Sidereal variations can arise 
in experiments with Earth-based sources
because the direction of neutrino propagation 
relative to the Sun-centered frame 
changes as the Earth rotates.
Annual variations can arise 
in solar-neutrino experiments
because the orientation of the detector relative to the Sun 
changes as the Earth orbits the Sun. 
{\it Compass asymmetries}
also result from rotation-symmetry breaking,
but the signals are independent of time.
They would appear at the location of the detector
as unexplained horizontal or vertical asymmetries. 
{\it Neutrino-antineutrino mixing}
is a direct consequence of any 
model with nonzero coefficients of type $g$ or $H$.
This class of signal involves lepton-number violation,
and it includes in particular any appearance measurement 
that can be traced to $\nu\mix\nub$ oscillation.
The final class of signal
involves the {\it classic CPT test} of the relation
$P_{\nu_b\to\nu_a}\ne P_{\nub_a\to\nub_b}$.
This could also include violation of the condition 
$P_{\nu_b\rightleftarrows\nub_a}(t) = 
P_{\nu_a\rightleftarrows\nub_b}(t)$,
which requires $\nu\mix\nub$ mixing.

Overall, 
the prospects are good for studies of Lorentz and CPT violation 
via neutrino oscillations. 
Consider an experiment
with maximum $L$ coverage of $L_{\rm max}$
and maximum $E$ coverage of $E_{\rm max}$.
A rough estimate of the sensitivity $\si$
to a coefficient for Lorentz violation of dimension $1-d$
can be taken as\cite{nu2}
$\si \approx - \log L_{\rm max} - d \log E_{\rm max}$.
This can be used to show that all classes of experiment
can attain Planck-scale reach for Lorentz and CPT violation,
and the best may rival some of the most sensitive experiments
in other sectors of the SME.
Note also that non-oscillation experiments with neutrinos 
typically also have sensitivity to Lorentz violation,\cite{nu2}
in particular via sidereal variations and compass asymmetries.
This includes laboratory experiments such as direct mass searches 
or searches for neutrinoless double-beta decay
and also astrophysical observations of supernova neutrinos.

\section{Illustrative Models}\label{models}

To gain insight into the variety of oscillation behavior exhibited
by neutrinos in the presence of Lorentz violation,
it is useful to consider special cases
of the effective hamiltonian \rf{heff}
involving only a few nonzero coefficients.
In this part of the talk,
some possible models of this type are first briefly discussed,
and then a realistic example (the bicycle model) is considered.

\subsection{Basics}

One class of special models
is obtained by requiring rotation symmetry.
The resulting rotation-invariant 
or so-called `fried-chicken' (FC) models
are of definite interest for certain investigations because 
the rotation symmetry simplifies some calculations.
It is worth noting,
however,
that these models are difficult to motivate theoretically
as exact descriptions relevant to experimental studies.
Thus,
it might seem reasonable to require spherical symmetry 
in a special frame,
perhaps the cosmic microwave background (CMB) frame.
However,
assuming rotation symmetry in the CMB frame
means that the coefficients appearing 
in the effective hamiltonian \rf{heff}
differ from those appearing in the standard Sun-centered frame
or any other experimentally attainable frame.
Converting between the CMB frame and 
the Sun-centered frame or other standard choice of 
experimentally attainable frame
introduces direction dependence 
due to the solar-system motion relative to the CMB. 
This implies that the experimentally relevant hamiltonian 
also involves spatial components of the coefficients,
suggesting that it cannot be an FC limit 
of the theory \rf{heff}. 

The general FC limit of the effective hamiltonian $\heff$ 
is straightforward to obtain.
It contains four matrices,
$(\mt)_{ab}$,
$(a_L)^0_{ab}$,
$(c_L)^{00}_{ab}$,
$(c_L)^{jk}_{ab}=\fr13(c_L)^{ll}_{ab}\de^{jk}$.
However,
only three of these are independent
because the trace $(c_L)^{00}_{ab}-(c_L)^{jj}_{ab}$ is
unobservable and can be set to zero.
For definiteness,
assume the rotation symmetry occurs
in the Sun-centered $(T,X,Y,Z)$ frame.
Then,
the relevant part of the $6\times6$ effective hamiltonian 
reduces to the block-diagonal form
\bea
(\heff)^{\rm FC}_{ab}&=&
\left(\begin{array}{cc}
\big( \mt/2E +(a_L)^T -\frac43(c_L)^{TT}E \big)_{ab} 
\hskip -20pt 
& 
0 \\
\hskip -20pt 
0 & 
\hskip -50pt
\big( \mt/2E -(a_L)^T -\frac43(c_L)^{TT}E \big)^*_{ab}
\end{array}\right) .
\label{rim}
\eea
This equation determines the general FC model
for three active neutrinos.
Additional light or massless sterile neutrinos 
can be incorporated if needed.

Except for the original proposal 
for Lorentz violation in neutrinos\cite{ck}
and more recent papers,\cite{nu,nu2,chki}
a sizable part of the literature\cite{fc1,fc2,fc3,fc4}
concerns restricted special limits of the general FC model \rf{rim}.
In particular, 
most works have considered the two-generation special case
in the limit of vanishing $(a_L)^T$ or $(c_L)^{TT}$.
Even though the FC model \rf{rim} is somewhat restricted
relative to the rich structure 
of the full effective hamiltonian \rf{heff},
a more complete study would be of definite interest.
A large class of oscillatory behaviors and their phenomenological
implications in the FC limit remain unexplored to date.

In the more general context
of the full effective hamiltonian \rf{heff},
Lorentz violation implies directional dependence 
of oscillation physics through the breaking of rotation invariance.
A variety of special cases involving directional dependence
can be considered.\cite{nu2}
For example,
one interesting limiting class of models with direction dependence
consists of those with nonzero coefficients
$g^{\mn\si}$ and $H^\mn$ only.
These models necessarily involve $\nu\mix\nub$ mixing.
In the general theory \rf{heff},
nonzero $\nu\mix\nub$ mixing of this type can lead 
to five distinct oscillation lengths
and corresponding complications.

A restriction of the effective hamiltonian \rf{heff}
to the special case 
involving only the two-dimensional $\nu_e$-$\nub_e$ subspace
offers an interesting and readily workable limit
with directional dependence.
In this single-flavor limit of $\nu\mix\nub$ mixing,
any coefficients $(\mt)_{ee}$ and $(c_L)_{ee}$ 
can be ignored because
they are real and lead to terms proportional to the identity
that have no effect on oscillatory behavior.
Furthermore,
the coefficient $(H^\mn\cmat)_{ab}$ 
is antisymmetric in generation space, 
so $(H^\mn\cmat)_{ee}=H^\mn_{ee^C}=0$.
The most general single-flavor theory without mass differences
is therefore given by a $2\times 2$ effective hamiltonian
containing only the coefficients
$(a_L)_{ee}^\mu$ and $(g^{\mn\si}\cmat)_{ee}=g^{\mn\si}_{ee^C}$
for Lorentz violation.
This model and any further restrictions necessarily violate CPT,
since both $(a_L)_{ee}^\mu$ and 
$g^{\mn\si}_{ee^C}$ control CPT-odd terms in the hamiltonian.
Note that the probabilities for this general single-flavor model
have structure identical 
to that of the standard two-generation mixing case,
$P_{\nu_e\mix\nub_e}=
1-P_{\nu_e\to\nu_e}=
1-P_{\nub_e\to\nub_e}
=\sin^22\theta\sin^2 2\pi L/L_0$.
However, 
the mixing angle and oscillation length 
depend on the 4-momentum in an unconventional 
and direction-dependent way.

The above discussion has focused on basic consequences 
of special models with only a small number of coefficients.
However,
even in apparently simple cases of these types, 
the neutrino behavior can be complicated and counterintuitive.
An example of a counterintuitive phenomenon is given 
by the Lorentz-violating seesaw mechanism.\cite{nu2}
This can arise when there are degeneracies in the
low- or high-energy limit of $\heff$.
The different energy dependences 
among the coefficients for Lorentz violation in $\heff$ 
can then lead to the emergence of an oscillation length
behaving like a mass-squared difference,
despite the absence of mass-squared differences 
in the model.

To illustrate this,
consider a $3\times 3$ effective hamiltonian $\heff$ 
parametrized in some basis as
\beq
\heff = \left(\begin{array}{ccc}
2h_1   & h_2 & h_3 \\
h^*_2 & 0 & 0   \\
h^*_3 & 0   & 0
\end{array}\right) ,
\eeq
where irrelevant diagonal terms have been disregarded.
For this class of toy models,
the interesting eigenvalue difference is
$\De = \sqrt{(h_1)^2+|h_2|^2+|h_3|^2}-h_1$.
If the combination coefficients for Lorentz violation 
is such that $h_1\gg\sqrt{|h_2|^2+|h_3|^2}$ 
holds at some energy scale,
then the eigenvalue difference is approximately
$\De \approx \half(|h_2|^2+|h_3|^2)/h_1+\cdots$.
For instance,
suppose $h_2$ and $h_3$ originate from a dimension-one coefficient 
and hence are constant with energy,
while $h_1$ originates from a dimensionless coefficient 
that grows linearly with energy.
Then,
at high energies the eigenvalue difference 
is proportional to $E^{-1}$,
just like a standard neutrino mass difference.
More generally,
using various combinations of masses 
and coefficients for Lorentz violation,
similar models can be found that yield 
$E^{-1}$, $E^{-2}$, or $E^{-3}$ dependence at high energies,
or $E^1$, $E^2$, or $E^3$ dependence at low energies.
Further $E^n$ dependences can be obtained  
if the full $6\times 6$ effective hamiltonian \rf{heff}
is used.

\subsection{Bicycle model}

The complexity of the effective hamiltonian \rf{heff}
and the range of possible neutrino behavior
suggests that existing neutrino oscillation data 
may be compatible with an origin in Lorentz violation
without mass differences.
This is demonstrated explicitly 
by a simple model,
called the bicycle model,\cite{nu}
reproducing the major features 
of known neutrino behavior.

The bicycle model 
is a two-coefficient three-generation special case
of the theory \rf{heff}
without either mass-squared differences or $\nu\mix\nub$ mixing.
It therefore involves only two degrees of freedom,
rather than the four degrees of freedom
used in the standard description with mass. 
The nonzero coefficients include
an isotropic $c_L$ with nonzero element 
$\fr43(c_L)^{TT}_{ee} \equiv 2\cee >0$
and an anisotropic $a_L$ with degenerate nonzero real elements 
$(a_L)^Z_{e\mu}=(a_L)^Z_{e\ta}\equiv\a3em/\sqrt{2}$.
These coefficients are understood to be specified in 
the standard Sun-centered frame $(T,X,Y,Z)$.

Diagonalizing the hamiltonian for the model yields\cite{nu} 
\bea
P_{\nu_e\to\nu_e}&=&
1-4\sin^2\theta\cos^2\theta\sin^2(\De_{31}L/2) ,
\nonumber\\
P_{\nu_e\mix\nu_\mu}
&=& P_{\nu_e\mix\nu_\ta} =2\sin^2\theta\cos^2\theta\sin^2(\De_{31}L/2) ,
\nonumber\\
P_{\nu_\mu\to\nu_\mu}
&=& P_{\nu_\ta\to\nu_\ta}
=1-\sin^2\theta\sin^2(\De_{21}L/2) 
\nonumber\\ &&
\pt{= P_{\nu_\ta\to\nu_\ta}}
-\sin^2\theta\cos^2\theta\sin^2(\De_{31}L/2) 
-\cos^2\theta\sin^2(\De_{32}L/2) , \nonumber \\
P_{\nu_\mu\mix\nu_\ta}&=&
\sin^2\theta\sin^2(\De_{21}L/2) 
\nonumber\\ &&
-\sin^2\theta\cos^2\theta\sin^2(\De_{31}L/2)
+\cos^2\theta\sin^2(\De_{32}L/2) ,
\eea
where
\bea
\De_{21}&=&\sqrt{(\cee E)^2+(\a3em\cos\Th)^2}+\cee E , 
\nonumber \\
\De_{31}&=&2\sqrt{(\cee E)^2+(\a3em\cos\Th)^2} ,
\nonumber \\
\De_{32}&=&\sqrt{(\cee E)^2+(\a3em\cos\Th)^2}-\cee E ,
\nonumber \\
\sin^2\theta &=&\half [1-{\cee E}/
{\sqrt{(\cee E)^2+(\a3em\cos\Th)^2}}] ,
\label{sinth}
\eea
and where the propagation direction $\hat p$ is defined 
in Eq.\ \rf{vectors}.
These probabilities also hold for antineutrinos.

Consider first the qualitative features of this theory.
At low energies the coefficient $\a3em$ induces oscillation 
of $\nu_e$ into an equal mixture of $\nu_\mu$ and $\nu_\ta$,
while at high energies
the coefficient $\cee$ controls the physics
and there is no $\nu_e$ mixing.
The critical energy for the theory is given by
$E_0=|\a3em |/\cee$.
At energies $E\gg E_0$, 
Eq.\ \rf{sinth} shows that $\sin^2\theta$ effectively vanishes. 
The probabilities become those 
of a two-generation model with maximal $\nu_\mu\mix\nu_\ta$ mixing
and transition probability
$P_{\nu_\mu\mix\nu_\ta}\simeq\sin^2(\De_{32}L/2)$,
$\De_{32}\simeq\a3em^2\cos^2\Th/2\cee E$. 
This theory therefore incorporates a Lorentz-violating seesaw
yielding a pseudomass at high energies,
having energy dependence like  
that of a conventional mass-squared difference 
$\Dtm=\Dtmz\cos^2\Th$, 
where $\Dtmz = \a3em^2/\cee$.
However,
the value of the pseudomass $\Dtm$ 
and therefore the behavior of neutrino oscillations
vary with the declination $\Th$ of the propagation.
High-energy neutrinos 
propagating in the equatorial plane
undergo no oscillation because 
all off-diagonal terms in $\heff$ vanish,
while those propagating parallel to celestial north or south
behave according to the maximum pseudomass $\Dtmz$.

The existing data for oscillations of atmospheric neutrinos\cite{pdg}  
are compatible with neutrino behavior in this model.
To be specific, 
suppose $\Dtmz=10^{-3}$ eV$^2$ and $E_0=0.1$ GeV.
This corresponds to $\cee=10^{-19}$, $\a3em=10^{-20}$ GeV,
values that are consistent with the notion of Planck-scale suppression.
The value of $\Dtmz$ lies near 
the accepted range adopted in the standard mass-based analysis,
and $E_0$ then lies well below the energies
relevant for atmospheric neutrino data.
With these choices,
high-energy atmospheric neutrinos exhibit 
the standard energy dependence
even though they have zero mass differences,
thanks to the Lorentz-violating seesaw.
Moreover,
within existing experimental resolution
the zenith-angle dependence of the probability 
$P_{\nu_\mu\to\nu_\mu}$
averaged over the azimuthal angle 
is also similar  
to the standard maximal-mixing case
with two generations and a mass-squared difference
$\De m^2=2\times10^{-3}$ eV$^2$.
Nonetheless,
atmospheric-neutrino signals for Lorentz violation exist
that are distinct from standard behaviors.
For example,
the theory predicts compass asymmetries,
including significant {\it azimuthal} dependences.
This can be seen by considering neutrinos 
propagating in the horizontal plane of the detector.
Neutrinos coming from the north or south
experience a pseudomass of $\Dtm=\Dtmz\cos^2\ch$
and hence oscillate,
while those coming from the east or west have $\cos\Th=0$,
$\Dtm=0$ and hence experience no oscillations.

The basic solar-neutrino behavior predicted by the bicycle model 
is also consistent with observational data.\cite{pdg}
Since the Earth's orbital plane lies at $\et\simeq23^\circ$ 
relative to the equatorial plane,
and since solar neutrinos observed at the Earth
are those propagating in the orbital plane,
the value of $\cos^2\Th$ varies during the year.
It is zero at the two equinoxes,
and it reaches a maximum of $\sin^223^\circ$ at the two solstices.
Under the simplifying assumption 
of adiabatic propagation in the Sun,
the average $\nu_e$ survival probability is 
$\vev{P_{\nu_e\to\nu_e}}_{\rm adiabatic}=
\sin^2\theta\sin^2\theta_0+\cos^2\theta\cos^2\theta_0$.
Here,
$\theta_0$ is the solar-core mixing angle,
determined from Eq.\ \rf{sinth}
via substitution of $-\cee E$ with $-\cee E + G_Fn_e/\sqrt{2}$.
It therefore follows that the predicted neutrino flux 
in the bicycle model is half the expected value at low energies
and decreases at higher energies.
This is consistent with existing data.
Note that care is needed in interpreting the details 
of the time-dependent behavior.
Although the adiabatic probability lies near the average
during most of the year,
the adiabatic approximation gives a strong suppression
near the equinoxes. 
However,
the model predicts that in fact oscillations cease at these times,
so the adiabatic approximation fails then 
and the actual predicted survival probability spikes instead. 
The net combination of these effects near the equinoxes
generates fluctuations in the binned flux,
which offer a potential signal for Lorentz violation
that could be sought in a detailed experimental analysis.
The reader is cautioned,
however,
that although detection of the semiannual fluctuations 
represent a positive signal for Lorentz violation,
the lack of such a signal is insufficient to exclude
the bicycle model because
simple changes to the model
can yield only small semiannual fluctuations
while maintaining the other neutrino behaviors 
discussed above.\cite{nu}

The predictions of the bicycle model are also consistent 
with neutrino-oscillation data from other experiments.\cite{pdg}
For instance,
$\nub_e$ survival is determined by 
the oscillation length $2\pi/\De_{31}$,
which is sufficiently small to affect KamLAND.\cite{kamland}
The average $\nub_e$ survival probability is 
$\vev{P_{\nub_e\to\nub_e}}
=1-2\sin^2\theta\cos^2\theta \ge 1/2$,
so the observed flux is predicted to be somewhat above
half the flux expected in the absence of oscillations,
as indeed reflected by the existing data.
Similarly,
tests of the bicycle model are also possible
in long-baseline accelerator-based experiments,
which involve $\nu_\mu$ oscillations 
at GeV energies
and baselines of hundreds of kilometers.
In particular,
sidereal variations in $\nu_\mu\mix\nu_\ta$ mixing 
are to be expected.
The model also predicts that the results will vary
with the beamline-direction of the experiment
because the propagation angle $\Th$ and hence
the pseudomass $\Dtm = \Dtmz\cos^2\Th$ differ. 

In summary,
the bicycle model demonstrates that it is difficult
and probably impossible at present 
to exclude the idea that observed neutrino oscillations 
originate from Lorentz and CPT violation 
rather than from mass differences. 
Positive and definitive signals for Lorentz and CPT violation
arising from Planck-scale physics 
might be first revealed by detailed analysis 
of existing or near-future experimental data.

This work was supported in part by the 
United States Department of Energy
under grant number DE-FG02-91ER40661
and by the 
National Aeronautics and Space Administration
under grant number NAG8-1770.

\end{document}